# Correction to Bjorken Energy Density Calculations for Central *A–A* Collisions

© 2023 O. M. Shaposhnikova[1)], A. A. Marova[2)], and G. A. Feofilov[2)*]

**Abstract** We present geometry-based corrections to estimates of the Bjorken energy density in a broad range of heavy-ion collision energies (from RHIC to the LHC) considereing non-circular shape of the nuclei overlap area in case of events selected in 0–5% centrality classes. We compare the updated Bjorken energy excitation function, obtained in our study for these very central *A–A* collisions, to the previously obtained ones. We present also and discuss the relevant energy dependences of pion, kaon and proton contributions to the Bjorken energy density that are also estimated in our study.

## 1. INTRODUCTION

The mean energy density ε that is reached in the collisions of high-energy heavy ions is usually estimated using the hydrodynamic model originally proposed by Bjorken [1]. From early numerical studies in lattice QCD [2], the critical energy density required for formation of the QGP — deconfined matter of quarks and gluons, is known to be on the order of 1 GeV/fm$^3$ [2]. Following [1], ε is defined using the transverse energy $\frac{dE_\perp}{dy}$ of particles produced at midrapidity in the cylinder volume with the cross sectional area $S_\perp$ determined by the overlap between the colliding nuclei and the length relevant to the characteristic particle formation time τ as:

$$\varepsilon = \frac{dE_\perp}{dy} \cdot \frac{1}{S_\perp \tau} \ . \tag{1}$$

---

[1)]St. Petersburg Governor's Physics and Mathematics Lyceum 30, Saint Petersburg, Russia.
[2)]Saint-Petersburg State University, Saint Petersburg, Russia.
*E-mail: g.feofilov@spbu.ru

The transverse energy density $\frac{dE_\perp}{dy}$ includes here the contribution of all types of particles produced[3]. Since the neutral particles' momenta $p_\perp$ are not measured, then in order to take into account the main contribution from pions, kaons, and protons, the following approximation is used for the estimate of the mean value of $\frac{dE_\perp}{dy}$ [7]:

$$\frac{d<E_\perp>}{dy} = \frac{3}{2}(<m_\perp> \frac{dN}{dy})_{\pi^\pm} + 2(<m_\perp> \frac{dN}{dy})_{K^\pm,p,p^-} \ . \tag{2}$$

Here, $<m_\perp>$ is calculated for particles with mass $m$ and transverse momentum $p_\perp$ as $<m_\perp> = \sqrt{m^2 + p_\perp^2}$. Factors 3/2 and 2 are taking into account the production of neutral particles.

As to the formation time $\tau$, there is no uniform approach because formula (1) diverges when $\tau \to 0$, and it could be taken differently in theoretical approaches (one may see discussions in [4, 7, 8–10]). Assuming that the formation time $\tau$ is ~1 fm/$c$, similar to other studies, we will consider not $\varepsilon$, but the product $\varepsilon \cdot \tau$:

$$\varepsilon \cdot \tau = \frac{dE_\perp}{dy} \cdot \frac{1}{S_\perp} \ . \tag{3}$$

The variable $S_\perp$ is of a certain theoretical interest. Besides $\varepsilon \cdot \tau$, $S_\perp$ is used to calculate such quantities as $\sqrt{\frac{dN/dy}{S_\perp}}$ (the square root of the hadron multiplicity per unit of rapidity and unit of $S_\perp$) and $<p_\perp>/\sqrt{\frac{dN/dy}{S_\perp}}$ (a ratio between the average transverse momentum and $\sqrt{\frac{dN/dy}{S_\perp}}$). Studies of dependence of these quantities on $\sqrt{S_{NN}}$ on $<p_\perp>$ and collision centrality are relevant and could bring constraints on the theoretical analysis, because the energy distribution of all the produced particles and the particle density are connected to the initial energy and entropy densities of the produced matter, one may see some discussions in [8–10].

---

[3]In case of pseudorapidity variable the transformation from rapidity [3] is given by Jacobian $J(y,\eta)$, so that a commonly used formula of energy density estimate is given by [4] as $\varepsilon = \frac{1}{S_\perp c \tau} J(y,\eta) \frac{dE_\perp}{d\eta}$. The average value of Jacobian $J(y,\eta) \sim 1.25$ at $\sqrt{S_{NN}} = 200$ GeV [5] and it is ~1.09 for central collisions at 2.76 TeV[6].

The value of overlap area $S_\perp$ of two colliding nuclei is closely related to the centrality of collision. However, neither the collision impact parameter $b$ nor the area $S_\perp$ can be measured directly in the experiment. Several approaches could be found to the estimates of the mean value of $S_\perp$ relevant to some class of events. The common approach is to select some observables related to definite classes of impact parameters by comparing them to those ones relevant to some phenomenological model of hadron collisiions. Amongst the proxies for collision centrality, the multiplicity of charged particles ($N_{ch}$) and number of nucleons-participants ($N_{part}$) are most usually applied. The $N_{ch}$ distributions could be directly measured and used for selection of centrality (multiplicity) classes, while $N_{part}$ could be also calculated along with the impact parameter $b$ using the Glauber Monte Carlo (GMC) approach [4, 5, 9–11].

GMC calculations are starting from the nuclear density profile of the colliding nuclei described by a well known Woods–Saxon distribution with parameters[4] of nucleus radius $R$ and diffusion $a$ (see also some data and references in Table 1). As discussed in detail in [12], the GMC is basing on the assumption that the quantity of interest, the impact parameter $b$ for colliding nuclei, is monotonically related to particle multiplicity values. The particles are produced in binary nucleon–nucleon collisions and their values of $N_{ch}$ depend on the inelastic nucleon–nucleon cross section $\sigma^{NN}_{inel}$, on the collision energy and on the geometry of the overlapping area $S_\perp$ of colliding nuclei. Besides, it is assumed that $S_\perp$ could be related to the number of participating nucleons ($N_{part}$) by fitting the measuered multipliicty density distribution $dN/d\eta$ with a so-called two-component model [17], where $dN/d\eta$ is proportional to $[(1 − x)N_{part} + xN_{coll}]$. This last assumption is aimed to balance, via free parameter $x$, the interplay of soft and hard processes of multiparticle production.

---

[4]Values of $R = 6.5$ fm and $a = 0.535$ fm were used for the Au nucleus [7]. Note that $R$ was increased in [7] from charged radius value 6.38 fm to 6.5 fm to approximate the effect of the neutron skin. Slightly larger value of $R = 6.62$ fm $a = 0.546$ fm is used for the Pb nucleus [10, 11].

Thus, the GMC approach gives the possibility to define the area $S_\perp$ by averaging, over many events, of the maximum values of the transverse plane coordinate projections of position distributions of the participating nucleons. Such calculations of $b$, $N_{\text{part}}$, $N_{ch}$, and $S_\perp$ bring finally a certain low-bound level of $\varepsilon\cdot\tau$ values, details could be found in [4, 5, 9–11]. As one may see, in case of the given approach, there is a rather large number of assumptions used in the application of the GMC to the estimate of the overlapping area $S_\perp$. A possible bias in calculations of $N_{\text{part}}$ values, obtained in different GMC analyses, may (under)overestimate the overlap area $S_\perp$ thus leading to relevant bias of $\varepsilon\cdot\tau$ values. One of the illustrative examples of the dramatic influence of some assumptions in GMC could be found in [13] where two different methods were used. Two extremes vere considered in [13] for the overlap transverse area spanned by the participating nucleons: i) the exclusive (or direct) overlap between participating nucleons, ii) the inclusive (or full) area of all participating nucleons. As a result, it was found that in case of very central classes of Pb–Pb collisions at $\sqrt{S_{NN}}$ = 5.02 TeV the lower-bound Bjorken energy density estimate can reach values between 10 and 20 GeV/(fm$^2c$) [13].

In our study we re-examine below the usual aproach to Bjorken energy density calculations for very central 0–5% classes' collisons. We make also the analytical aproximation of the updated Bjorken energy excitation function, obtained in our study for these very central Au+Au and Pb+Pb collisions. Finally, we present and briefly discuss the relevant energy dependences of the Bjorken energy density of pion, kaon, and proton contributions that are also estimated in our study.

## 2. METHODOLOGY AND RESULTS

Instead of applying a standard way of GMC fitting of $N_{ch}$ disstributions in minimum bias Au–Au or Pb–Pb collisions, selection of a class of 0–5% central events, estimates of relevant number of participating nucleons $N_{\text{part}}$ and estimates of $S_\perp$, we use a simplified initial geometry of collision approach. Pure geometrical calculations, assuming collliding disks of radii $R = 6.87$ fm [15] for Au–Au collisions (or $R = 7.17$ fm [16] in Pb–Pb collisions), were performed for the case of 0–5% centrality class of impact parameter $b$. Results show that in both cases the dominating events, selected in the given class, will have a noticeable shift of $b$ from 0. Namely, the following values of mean impact parameter were obtained: $<b> = 2.18$ fm (for 0–5% class of Au–Au collisiins) and $<b> = 2.26$ fm (in case of Pb–Pb). Naturally, this non-zero value of the mean impact parameter, relevant to the given 0–5% centrality class, will result in a smaller overlap area of two nuclei. The GMC calculations for Au+Au collisions at three RHIC energies gave similar results $<b> \sim 2.2, 2.39, 2.21$ fm, see in [7]. And ~2.4 fm can be observed in Pb+Pb 0–5% centrality class of collisions at the LHC [11].

In our work, first of all, we obtained the estimate of $<\frac{dE_\perp}{dy}>$ by formula (2), taking into account the contribution of pions, kaons, and protons. For the values of $<m_\perp>$ and $(d<E_\perp>)/dy$, the published data on $<dN/dy>$ and $<p>$ were used (see references in Table 1 and Table 2). Our results for $<\frac{dE_\perp}{dy}>$ have been verified with significant data on $\varepsilon\cdot\tau$: for example, for an energy of 200 GeV the value $\varepsilon\cdot\tau = 5.07 \pm 0.4$ GeV/fm$^2$ agrees within the error with the data from [7]. These results are then used to calculate the product of energy density and formation time ($\varepsilon\cdot\tau$) presented in Fig. 1 and of particle contributions to this value using our modified estimates of $S_\perp$ shown in Fig. 2. The values are summarised in Table 1 and Table 2.

In our approach, we assume that the measured particle production and transverse energy flow at midrapidity are relevant, in a given event of collision, to the initial nuclear geometry. The last one is taken for the given class of central 0–5% events from the corresponding mean area of two overlapping disks of given radii. If we denote as $a$, the value of 1/2 of the mean shift $<b>$ from 0 of the dominating mean impact parameter for the given central 0–5% class, we may obtain for the corrected overlap area $S_{\perp\text{corr}}$:

$$S_{\perp \text{corr}} = 2 \cdot \left( \frac{2\pi R^2 \cdot \arccos\frac{a}{R}}{360°} - a \cdot \sqrt{R^2 - a^2} \right) \ . \tag{4}$$

Note, that formula (4) is valid for the impact parameter values $b < 2R$ and here $a = <b>/2$. We used the disk (or nuclear core) radius taken at 90% density in the papers [14, 15]. Calculated values of the overlap area $S_{\perp \text{corr}}$ are presented in Table 1. We used the core radii taken at 90% level of nuclear density tadial distribution from the papers [14, 15]. As an example: earlier the value $S_\perp =148.03$ fm² was obtained for 0–5% class of Au–Au collisions at the energy of 200 GeV [7], while the relevant new value at this energy $S_{\perp \text{corr}}$ is ~118.8 fm² (see Table 1). Therefore, the decrease of mean $S_\perp$ will result in the relevant increase of the value $\varepsilon \cdot \tau$. Results of recalculated values of $\varepsilon \cdot \tau$ for RHIC and LHC energies are shown in Table 1 and in Fig. 1. One may see a general increase in values of $\varepsilon \cdot \tau$ for all data points. Assuming formation time $\tau \sim 1$ fm/$c$, we obtain the energy density at top energy of Pb–Pb collisions $\varepsilon = (17.9 \pm 0.5)$ GeV/$c^3$, that is close to the $20.9 \pm 0.5$ GeV/$c^3$ upper value from [13].

It was established previously [6] by the CMS at the LHC that the transverse energy density at midrapidity grows with $\sqrt{S_{NN}}$ more rapidly compared to lower energy data, where logarithmic scaling with $\sqrt{S_{NN}}$ was observed. We see a similar tendency here in Fig. 1 for values of $\varepsilon \cdot \tau$ vs. $\sqrt{S_{NN}}$ up to 200 GeV. Since the extrapolation of the Bjorken energy density at lower energies may not be entirely accurate, we did not continue it now, but results of this work indicate that the critical energy density value of 1 GeV/fm³ could be reached at lower collision energies than previously predicted.

To describe data in Fig. 1 numerically and to get the best extrapolation method, we made fits to the resulting dependence of $\varepsilon \cdot \tau$ on $\sqrt{S_{NN}}$ using power-law and logarithmic fits:

$$\varepsilon \cdot \tau = Q \cdot (S_{NN})^n \ , \tag{5}$$

$$\varepsilon \cdot \tau = A + B \cdot \ln(\sqrt{S_{NN}}) \ . \tag{6}$$

Results of fitting the excitation functions for ε·τ are shown in Fig. 1 and Fig. 2. The values of parameters are summarized in Table 3.

Also, in this paper, we considered the contributions of particle yields to the Bjorken energy density (see Fig. 2). We made our calculations of $<\frac{dE_\perp}{dy}>$ for pions, kaons, and protons using the data from [7, 11, 18] on $<\frac{dN}{dy}>$ and $<p_\perp>$. Results of calculations of the value ε·τ are presented in Fig. 2. Fitting parameters are presented in Table 4.

One may see that pions and kaons make the greatest contribution, but protons cannot be neglected either in the general balance of Bjorken energy. Also, it can be seen that heavy particles have a slower dependence of product ε·τ on $\sqrt{S_{NN}}$. One may compare the result of power law fitting parameter $n = 0.162$ obtained in fitting of these different contributions, it is very close to the one obtained for aproximation of energy dependence of particle multiplicuty in A–A collisions for $N_{ch}$ vs. $\sqrt{S_{NN}}$ ($n = 0.15$) [19]. Note that in case of *pp* collisions this parameter has the lower value % $n = 0.11$. One may assume that in the collisions at LHC the incident energy of colliding nuclei is converted mainly into production of particles at midrapidity rather than into increasing the particle transverse mass. These results are of a certain interest for further studies of particle production mechanisms in hadron collisions at the LHC.

## 3. CONCLUSIONS

For the most central 0–5% class of Au–Au and Pb–Pb collisions, we take into account the dominance of events with an average impact parameter $<b> \sim 2.2$ fm. In this regard, the collision overlap area decreases and, consequently, the value of Bjorken energy density estimate increases. The results of the recalculation of values of product of Bjorken energy density and formation time (ε·τ) are presented. Two approximations of the dependence of ε·τ on $\sqrt{S_{NN}}$ are checked and power law fit was found as the most apropriate in the whole energy range from RHIC to LHC.

In case of pions, the faster growth $\sqrt{S_{NN}}$ of the contribution to Bjorken energy density is observed compared to that for kaons and protons. But the contribution by protons to the general balance of Bjorken energy is also noticeable.

It can be expected from these results that the Bjorken critical energy region of ~1 GeV/fm$^3$ could be reached at lower *A–A* collision energy.

## ACNOWLEGMENTS

This work was supported by St. Petersburg State University project ID: 94031112.

FIGURE CAPTIONS

**Fig. 1.** Values of ε·τ as a function of $\sqrt{S_{NN}}$ for the most central 0–5% class of events at LHC energies. Experimental points (black circles): at 62.4 GeV, 130 GeV, 200 GeV [7], 2760 GeV [3], 5020 GeV [4]. Recalculated points using formula (4) (open circles). Lines: fitting with functions formulas (5) and (6) (see parameters in Table 2).

**Fig. 2.** Calculated contributions to ε·τ vs. $\sqrt{S_{NN}}$ using data [7, 11, 18] for pions, kaons, and protons, for the most central class of events at LHC energies.

**Table 1.** Nuclear radii, mean $E_\perp$ per unit rapidity, $S_\perp$, $S_{\perp\text{corr}}$ and $\varepsilon \cdot \tau$ in 0–5% central Au–Au and Pb–Pb collisions (we believe that the systematic density error due to different ways of trimming the diffuse edge is below 10%)

| | | | | | This work | |
|---|---|---|---|---|---|---|
| $\sqrt{S_{NN}}$, GeV | $S_\perp$, fm$^2$ | $\varepsilon \cdot \tau$, GeV/ fm$^2$ | $\frac{dE_\perp}{dy}$, GeV | $R$, fm | $S_{\perp\text{corr}}$, fm$^2$ | $\varepsilon \cdot \tau$, GeV/ fm$^2$ |
| 62.4 | 148 [14] | 3.7 ± 0.3 [7] | 524.01 | 6.87 [14] | 118.45 | 4.4 ± 0.3 |
| 130 | 148 [14] | 4.4 ± 0.3 [7] | 632.55 | 6.87 [14] | 118.45 | 5.3 ± 0.3 |
| 200 | 148 [14] | 5.2 ± 0.4 [7] | 750.34 | 6.87 [14] | 118.45 | 6.3 ± 0.4 |
| 2760 | 162.5 [15] | 12.3 ± 1.0 [12] | 1836.18 | 7.17 [14] | 129.46 | 14.2 ± 1.0 |
| 5020 | 162.5 [15] | 14.85 ± 0.53, see estimate in [16]; 10–20 [13] | 2318.94 | 7.17 [15] | 129.46 | 17.9 ± 0.5 |

**Table 2.** Contribution of pions, kaons, and protons to the mean $E_\perp$ per unit rapidity and to the product $\varepsilon \cdot \tau$ in 0–5% central Au–Au and Pb–Pb collisions

| $\sqrt{S_{NN}}$, GeV | Particles | $<p_\perp>$, GeV/$c$ | | $<\frac{dN}{dy}>$ | | $<\frac{dE_\perp}{dy}>$, GeV  This work | $\varepsilon \cdot \tau$, GeV/fm$^2$  This work |
|---|---|---|---|---|---|---|---|
| | | | | $\pi^+$, K+, p | $\pi^-$, K$^-$, $\overline{p}$ | | |
| 62.4 | Pions | 0.4 [7] | | 237 ± 17 [7] | 233 ± 17 [7] | 298.7 ± 14.3 | 2.5 ± 0.14 |
| | Kaons | 0.6-0.65 ± 0.05 [7] | | 32.4 ± 2.3 [7] | 37.6 ± 2.7 [7] | 111.5 ± 16.2 | 0.94 ± 0.16 |
| | Protons | 0.95 ± 01 [7] | | 13.6 ± 1.7 [7] | 29.0 ± 3.8 [7] | 113.8 ± 26.8 | 0.96 ± 0.27 |
| 130 | Pions | 0.4 [7] | | 280 ± 25 [[7] | 278 ± 25 [7] | 355 ± 21 | 2.99 ± 0.2 |
| | Kaons | 0.65-0.7 ± 0.05 [7] | | 42.7 ± 6.2 [7] | 46.3 ± 6.5 [7] | 145.8 ± 35.3 | 1.2 ± 0.35 |
| | Protons | 1 ± 01 [7] | | 20.0 ± 3.4 [7] | 28.2 ± 4.4 [7] | 132.2 ± 35.4 | 1.1 ± 0.35 |
| 200 | Pions | 0.4 [7] | | 327 ± 25 [7] | 322 ± 25 [7] | 392 ± 24 | 3.3 ± 0.23 |
| | Kaons | 0.7-0.8 ± 0.05 [7] | | 49.5 ± 6.2 [7] | 51.3 ± 6.5 [7] | 181 ± 40.8 | 1.5 ± 0.4 |
| | Protons | 1.1 ± 01 [7] | | 26.7 ± 3.4 [7] | 34.7 ± 4.4 [7] | 177.6 ± 41.3 | 1.5 ± 0.4 |
| 2760 | Pions | 0.517+-0.019 [11] | 0.520+-0.018 [11] | 733 ± 54 [11] | 732 ± 52 [11] | 1179.96 ± 110.7 | 9.1 ± 1.0 |
| | Kaons | 0.876+-0.026 [11] | 0.867+-0.027 [11] | 109 ± 9 [11] | 109 ± 9 [11] | 436.7 ± 56.5 | 3.4 ± 0.5 |
| | Protons | 1.333+-0.033 [11] | 1.353+-0.034 [11] | 34 ± 3 [11] | 33 ± 3 [11] | 219.5 ± 27.1 | 1.7 ± 0.24 |
| 5020 | Pions | 0.5682 [18] | | 1699.80 [18] | | 1491.8 ± 167.2 | 11.5 ± 1.5 |
| | Kaons | 0.9177 [18] | | 273.41 [18] | | 569.8 ± 34.8 | 4.4 ± 0.3 |
| | Protons | 1.4482 [18] | | 74.56 [18] | | 257 ± 18 | 1.99 ± 0.16 |

**Table 3.** Approximation parameters for values of ε · τ depending on $\sqrt{S_{NN}}$ for the most central 0–5% class of events from RHIC to LHC energies using formulas (5) and (6)

|  | Formula (7) | | | Formula (8) | |
| --- | --- | --- | --- | --- | --- |
|  | Q | n | χ²/NDF | A | B |
| This work | 1.12 ± 0.1 | 0.162 ± 0.004 | 0.23/3 | -2.3 ± 1.3 | 1.60 ± 0.27 |
| Experimental data of Fig. 1 | 0.932 ± 0.05 | 0.163 ± 0.003 | 0.085/3 | -- | |

**Table 4**. Approximation parameters for contributions of pions, kaons, and protons to $\varepsilon \cdot \tau$ versus $\sqrt{S_{NN}}$ for the most central 0–5% class of events from RHIC to LHC energies using formula (5) to fit data of Fig. 2

|  | Q | $n$ | $\chi^2$/NDF |
|---|---|---|---|
| Pions | 0.49 ± 0.04 | 0.185 ± 0.005 | 0.107/3 |
| Kaons | 0.23 ± 0.03 | 0.170 ± 0.008 | 0.036/3 |
| Protons | 0.61 ± 0.13 | 0.068 ± 0.015 | 0.088/3 |

Fig. 1

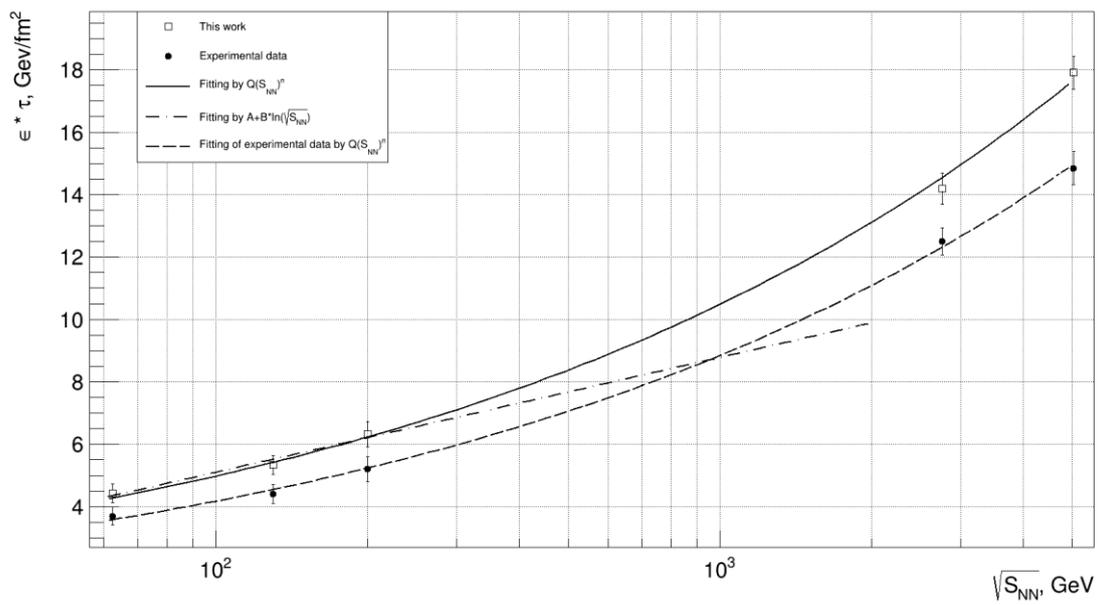

Fig 2.

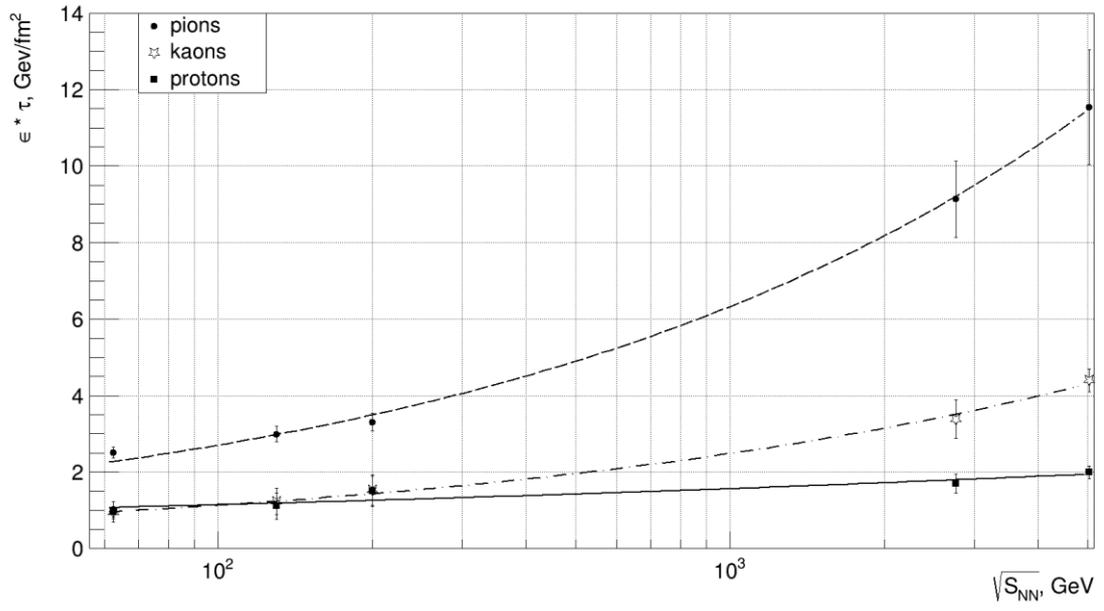